\documentclass[12pt]{article}
\usepackage[utf8]{inputenc}
\usepackage[margin=1in]{geometry}

\usepackage{graphics,graphicx}
\usepackage{amssymb}
\usepackage{natbib}
\usepackage{aas_macros}
\usepackage{enumitem}

\usepackage{titlesec}
\usepackage[breaklinks,colorlinks,urlcolor=blue,citecolor=blue,linkcolor=blue]{hyperref}
\usepackage[all]{hypcap}

\usepackage[dvipsnames]{xcolor}

\definecolor{myblue1}{RGB}{22, 63, 130}
\definecolor{mygrey1}{RGB}{145, 145, 145}

\usepackage{times}
\usepackage[T1]{fontenc}

\titlespacing\section{0pt}{6pt plus 4pt minus 2pt}{4pt plus 2pt minus 2pt}
\titlespacing\subsection{0pt}{6pt plus 4pt minus 2pt}{0pt plus 2pt minus 2pt}
\titlespacing\subsubsection{0pt}{6pt plus 4pt minus 2pt}{0pt plus 2pt minus 2pt}
\titlespacing\paragraph{0pt}{6pt plus 4pt minus 2pt}{12pt plus 2pt minus 2pt}

\titleformat*{\section}{\large\bfseries} 
\titleformat*{\subsection}{\normalsize\bfseries}
\titleformat*{\subsubsection}{\normalsize\bfseries}
\titleformat*{\paragraph}{\bfseries\color{myblue1}}

\begin{document}

\setlist[itemize]{label={},leftmargin=7mm,rightmargin=7mm}
\Large
\noindent Astro2020 APC White Paper: State of the Profession Consideration\\

\noindent The Early Career Perspective on the Coming Decade, Astrophysics Career Paths, and the Decadal Survey Process \\

\normalsize
\vspace{1mm}

\noindent \textbf{Thematic Areas:} State of the profession considerations: Early career concerns for the coming decade, graduate and postdoctoral training, career preparation, career transitions, and structure and dissemination of decadal survey. \\

\noindent\textbf{Principal Authors:}

\noindent Names: Emily Moravec | Ian Czekala | Kate Follette	

\noindent Institutions: University of Florida, 2018 NASEM Christine Mirzayan Science and Technology Policy Fellow | UC Berkeley | Amherst College  

\noindent Emails: emoravec@ufl.edu | iancze@gmail.com | kfollette@amherst.edu \\


\noindent \textbf{Co-authors:} Zeeshan Ahmed, Mehmet Alpaslan, Alexandra Amon, Will Armentrout, Giada Arney, Darcy Barron, Eric Bellm, Amy Bender, Joanna Bridge, Knicole Colon, Rahul Datta, Casey DeRoo, Wanda Feng, Michael Florian, Travis Gabriel, Kirsten Hall, Erika Hamden, Nimish Hathi, Keith Hawkins, Keri Hoadley, Rebecca Jensen-Clem, Melodie Kao, Erin Kara, Kirit Karkare, Alina Kiessling, Amy Kimball, Allison Kirkpatrick, Paul La Plante, Jarron Leisenring, Miao Li, Jamie Lomax, Michael B. Lund, Jacqueline McCleary, Elisabeth Mills, Edward Montiel, Nicholas Nelson, Rebecca Nevin, Ryan Norris, Michelle Ntampaka, Christine O'Donnell, Eliad Peretz, Andres Plazas Malagon, Chanda Prescod-Weinstein, Anthony Pullen, Jared Rice, Rachael Roettenbacher, Robyn Sanderson, Jospeh Simon, Krista Lynne Smith, Kevin Stevenson, Todd Veach, Andrew Wetzel, and Allison Youngblood

\begin{center}
\includegraphics[width=\textwidth]{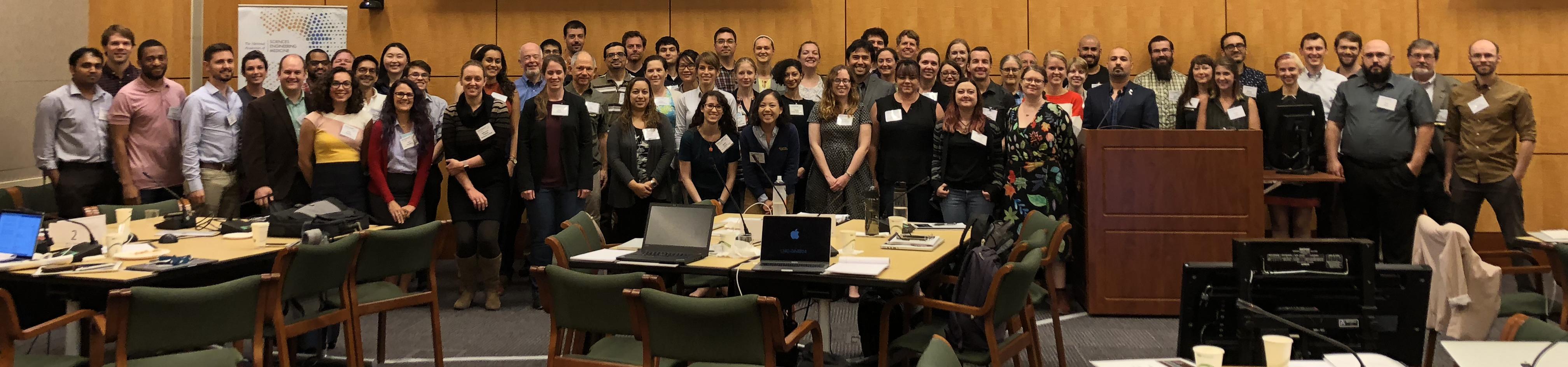}
\end{center}

\pagebreak

\noindent \textbf{Executive Summary:} In response to the need for the Astro2020 Decadal Survey to explicitly engage early career astronomers, the National Academies of Sciences, Engineering, and Medicine hosted the Early Career Astronomer and Astrophysicist Focus Session (ECFS)\footnote{http://sites.nationalacademies.org/SSB/SSB\_185166} on October 8-9, 2018 under the auspices of Committee of Astronomy and Astrophysics. The meeting was attended by fifty six pre-tenure faculty, research scientists, postdoctoral scholars, and senior graduate students\footnote{List of participants http://sites.nationalacademies.org/cs/groups/ssbsite/documents/webpage/ssb\_189919.pdf}, as well as eight former decadal survey committee members, who acted as facilitators. The event was designed to educate early career astronomers about the decadal survey process, to provide them with guidance toward writing effective white papers, to solicit their feedback on the role that early career astronomers should play in Astro2020, and to provide a forum for the discussion of a wide range of topics regarding the astrophysics career path. 

This white paper presents highlights and themes that emerged during two days of discussion. In Section 1, we discuss concerns that emerged regarding the coming decade and the astrophysics career path, as well as specific recommendations from participants regarding how to address them. We have organized these concerns and suggestions into five broad themes. These include (sequentially): (1) adequately training astronomers in the statistical and computational techniques necessary in an era of ``big data", (2) responses to the growth of collaborations and telescopes, (3) concerns about the adequacy of graduate and postdoctoral training, (4) the need for improvements in equity and inclusion in astronomy, and (5) smoothing and facilitating transitions between early career stages. Section 2 is focused on ideas regarding the decadal survey itself, including: incorporating early career voices, ensuring diverse input from a variety of stakeholders, and successfully and broadly disseminating the results of the survey.  

Recommendations presented here do not necessarily represent a universal consensus among participants, nor do they reflect the entirety of the discussions. Rather, we have endeavored to highlight themes, patterns, and concrete suggestions.  

\clearpage

\Large
\begin{centering}
\textbf{The Early Career Perspective on the Coming Decade, Astrophysics Career Paths, and the Decadal Survey Process}
\end{centering}
\\

\normalsize
\section{Discussion Themes: Topics of Importance for Early Career Professionals in the Coming Decade}
\subsection{Theme 1 - The big role of big data and data science in the 2020s}
Big data and data science have become integral parts of astronomical research. In an era of data proliferation (both observational and simulated), professional and funding agencies must consider the provision of fiscal, physical, and computational resources when planning projects and evaluating proposals. Particular needs in the area of data science include: the development of infrastructure to manage large datasets, the training of early career professionals in the techniques necessary to work with sophisticated data analysis tools, and the expansion and development of career paths focused on software development, data collection, and data analysis. We recommend that management, development, and analysis of big data be considered both within its own decadal panel and holistically across the survey panels and committees.

During the ECFS, ``big data" and modern data analysis techniques featured prominently in discussions about preparation for academic career paths, as well as transitions into professions that employ astronomy Ph.D.s (data science, engineering, defense). There was a general concern among participants that graduate curricula do not provide sufficient training in the sophisticated computational and statistical methodologies necessary to efficiently handle data in general, and large datasets in particular. Participants felt strongly that in order to incentivize the improvement of graduate and postdoctoral training in this area, the decadal survey committee should make explicit recommendations to funding agencies, professional societies, and departments regarding improved training.   

\paragraph{Recommendations}
\begin{itemize}
    \item \textbf{1.1.1} The decadal survey should include explicit prioritization of theoretical and computational questions in astronomy. 
    \item \textbf{1.1.2} The decadal survey should consider whether the evolution of the profession means that a redistribution or re-prioritization of permanent positions is necessary. In particular, ECFS participants suggested that the panels consider ways in which grant funding might be restructured to incentivize semi-permanent software development and data management positions rather than postdoctoral positions (see also Theme 5).
    \item \textbf{1.1.3} Professional societies and funding agencies should provide support for conference workshops on data science. Wherever possible, these should be embedded rather than pre-conference workshops to minimize the additional cost of attendance for early career participants.
    \item \textbf{1.1.4} Departments should restructure graduate curricula to explicitly address the computational and statistical techniques necessary to succeed in the big data era (see also Theme 3).
    \item \textbf{1.1.5} Observatories, national labs, and large collaborations (LSST, DES, etc.) should increase the number of data science workshops and data-science themed postdoctoral fellowships.
\end{itemize}

\subsection{Theme 2 - The evolving challenges of large missions and collaborations}
A further concern among ECFS participants was the trend toward larger facilities and projects. While acknowledging that flagship missions and thirty meter class telescopes are critical for advancing the field, early career astronomers voiced a need for the astronomical community to maintain a broad portfolio. There was widespread concern about the closing of current facilities to ``make room" in budgets for large telescopes and missions. Continued operation of an array of facilities operating in various wavelength regimes will be key for the professional training of future astronomers, as well as being needed more broadly for followup observations, time-domain monitoring, etc. We urge the decadal survey committee to consider the need for smaller and more easily accessible facilities that will allow early career astronomers to lead proposals, execute observations, and obtain their own data.    

While many early career astronomers are enthusiastic about the science enabled by large facilities and big collaborations, discussions at the ECFS revealed significant anxiety around the preparedness of Ph.D. astronomers to work in an era where `big' is the norm. Particular concerns in this regard are: the availability of a sufficient number of desirable long-term (non-postdoctoral) positions within academia to support large projects, the marketability of PhD astronomers in fields outside of or peripheral to academic astronomy, and concerns regarding whether and how contributions to large collaborations will be recognized by potential employers. The group predicted that more support scientist positions will be needed to design and maintain hardware and software for large facilities, but have observed a lack of sufficient funding for facility operations and maintenance under the current funding paradigm. 

\paragraph{Recommendations}
\begin{itemize}
\item \textbf{1.2.1} The decadal survey committee, funding agencies, and observatories should carefully consider the prioritization of and funding for construction of new facilities vs. operation and maintenance of existing facilities. A broad portfolio of facilities should be maintained, including those where early career astronomers can reasonably expect to lead proposals and obtain telescope time for smaller projects. 
\item \textbf{1.2.2} The reasonableness of cost estimates for missions and facilities proposed to the decadal survey should be carefully evaluated during project prioritization so that smaller facilities and support staff positions are not ``squeezed out" in agency budgets if and when the cost for these missions balloons.
\item \textbf{1.2.3} More astronomical journals should move toward a model where contributions to published papers are detailed specifically. This will incentivize early career astronomers to make contributions to large projects even when it will not lead to first authorship, and will provide a means for potential employers to gauge the role of applicants in coauthored papers.
\item \textbf{1.2.4} Professional societies and publishers should provide more opportunities for software development work to be recognized, published, and cited.
\end{itemize}

\subsection{Theme 3 - Career Preparation and Opportunities}
Early career astronomers are ideally situated to reflect on the experience of their education as well as assess its adequacy in preparing them for the challenge of navigating early career transitions. During the ECFS, participants focused on evaluating existing training, discussing the challenges that mark the transitions to postdoctoral and long-term positions, and preparing recommendations to improve these areas.

ECFS participants felt that some substantive changes are necessary to make the process of obtaining a Ph.D. more inclusive and supportive, and to reduce imbalances that lead to systemic inequalities in the profession (see Theme 5). In this section, we focus on recommendations around maximizing flexibility and marketability of Ph.D. astronomers, and around restructuring graduate curricula to reflect the changing nature of the profession. 

\paragraph{Recommendations}
\begin{itemize}
\item \textbf{1.3.1} Departments should incorporate ethics and professional development seminars into graduate curricula. In addition to focusing on proper scientific and professional conduct, these seminars are an excellent venue to hone research and presentation skills.
\item \textbf{1.3.2} Departments should consider the flexibility of their curricula, de-emphasizing a burdensome number of core classes and allowing for more electives, particularly in statistical and computational methods.
\item \textbf{1.3.3} Grant components such as ``Mentoring Plans,'' and ``Facilities and Resources'' documents should be given more weight and structure. In particular, they should be tied specifically to curricular methodologies and professional development opportunities for all grants requesting graduate student or postdoctoral funding, and should be standardized across agencies.
\item \textbf{1.3.4} AAS should incentivize (e.g. through reduced fees for registration, exhibit space, and workshop facilitation) the participation of members of industry in annual conferences and should consider organizing an alternative careers networking event. 
\end{itemize}

\subsection{Theme 4 - The Postdoctoral Years}
While the postdoctoral scholar period can be a time of great opportunity and scholarly freedom, it is also fraught with uncertainty. The traditionally transient nature of the postdoctoral years, where frequent cross-country (or international) moves are common and even encouraged, has a variety of hidden societal costs. 
``2-body problems'' and relocation costs add stress and uncertainty to the career transition process and disproportionately affect women and underrepresented minorities.\footnote{See \url{http://womeninastronomy.blogspot.com/2013/02/figure-1-two-body-problem.html} and references therein.}
Efforts are also needed to de-stigmatize the actions of those astronomers who express interest in non-academic career options, including during the graduate admission process and during graduate and postdoctoral mentoring.  
\paragraph{Recommendations} 
\begin{itemize}
\item  \textbf{1.4.1} Five year postdoctoral appointments should be encouraged by funding agencies and appointments with terms of less than three years should be strongly discouraged.
\item \textbf{1.4.2} To increase the quality of applications while also reducing the considerable stress on applicants during job application season, the postdoctoral fellowship application process should be standardized in terms of the nature and length of required materials. Specific and measurable criteria should be posted in job ads to help applicants in evaluating their suitability for positions. For example, vague advertising for ``outstanding candidates'' preselects for candidates with high self-efficacy.
\item \textbf{1.4.3} The AAS should contract with social scientists to conduct regular membership surveys that include astronomers who have left academia, and should include the results in state of the profession reports.  
\item \textbf{1.4.4} The social and psychological pressures of the Astrophysics Jobs Rumor Mill outweigh the benefits, and its information is often inaccurate or incomplete. It should be replaced by a commitment on the part of departments and agencies offering employment to release timely information to \textbf{all} job applicants regarding the search status, even if it has not concluded (e.g. ``we have created a short list and invited five applicants to visit").
\item \textbf{1.4.5} Development of teaching and communication skills should be incentivized and given greater emphasis in graduate curricula, postdoctoral training, and analysis of candidates for academic jobs. This can be accomplished through targeted graduate coursework, workshops, structured feedback to graduate students giving presentations (i.e. journal clubs), a greater emphasis on teaching and diversity statements in job applications, etc.  
\item \textbf{1.4.6} The community should rethink the distribution of temporary (e.g. postdoc) vs. potentially permanent (e.g. staff scientist) positions and the ways in which they are funded (e.g. grants vs. operations budgets). In particular, support and software design for large surveys and telescopes should be done by long-term staff members and not by postdocs. 
\end{itemize}

\subsection{Theme 5 - A Diverse Workforce and Equitable Community}
There was a strong consensus and desire among attendees that the astronomy community of the future should reflect the breadth of identities of the nation as a whole. Progress has been made in this area in the past decade, however there remains a great deal of work to be done. In particular, more work is needed to eliminate inequities and barriers to entry into the profession, which will enable more people of color, white women, members of the LGBTQ+ community, and people with disabilities to enter and, more importantly, to remain in the field. These efforts will benefit \textbf{all} members of the profession in myriad ways. In order to enable meaningful change, we hope to see the 2020 decadal survey make specific, high-priority, and far-reaching recommendations in this regard. 

\paragraph{Recommendations}
\begin{itemize}
\item \textbf{1.5.1} We urge the decadal panels to consider white papers regarding the state of the profession to be of equal importance to science and facilities white papers. Although these papers will be read by separate panels, we hope that members of the state of the profession panels will work together with members of the science panels and survey officials to ensure that the broader recommendations of Astro2020 integrate science and state of the profession recommendations powerfully and effectively. 
\item \textbf{1.5.2} Minoritized individuals often bear a disproportionate service burden, including during graduate school and postdoc years when these efforts are not generally a recognized responsibility. Graduate and postdoctoral fellowships should explicitly ask for information about, value, and reward this important work.
\item \textbf{1.5.3} AAS should consider establishing an award that specifically recognizes service in the area of improving representation in our profession.
\item \textbf{1.5.4} Training concerning gender and racial harassment should be required for all members of academic departments, including: faculty, postdocs, graduate students (especially teaching assistants), and staff. More training is needed across the board to combat gender and racial harassment in our field, particularly training with a focus on intersectionality.  
\item \textbf{1.5.5} To mitigate concerns about harassment and bullying, graduate departments should move away from the single mentor model towards advising strategies that encourage interaction with multiple faculty members on a regular basis (e.g., larger advising committees with semi-regular meetings). This would have the added benefit of proving more opportunities for all students to seek support and advice from a range of mentors in addition to providing a safety net for students with difficult or abusive advising relationships.
\item \textbf{1.5.6} The AAS should establish a list of best-practices for parental and child leave policies for graduate students and postdocs. Similar to recommendation 1.3.3, this information should be included in grant documents that request funding for graduate students and postdocs.
\end{itemize}

\section{Recommendations Regarding the Structure and Dissemination of the Decadal Survey}
\subsection{Early career participation in the decadal survey}
A key goal of the ECFS was to solicit feedback regarding how early career astronomers would like to be represented throughout the decadal survey process. Many participants argued for the most direct form of representation: full membership on survey panels. Several concerns regarding full panel membership were raised, however. These included: 1) reduced scientific productivity during a vulnerable career stage, 2) uncertainty regarding recognition by tenure committees and potential employers of the importance of this service to the profession, and 3) the potential to become enmeshed in controversial decisions that could negatively affect future career prospects. Several ideas were put forward regarding ways to mitigate these concerns for early career participants on the panels, namely: funding (e.g., a ``decadal fellowship''), recognition (e.g. a specific title or award that could be included in job and tenure applications), commitments from panel chairs to write letters of recommendation for jobs and letters of support for tenure, commitments from senior panel members to support early career colleagues with invitations to give seminars and colloquia, and childcare support.

The principal alternative to full participation put forward was the formation of ``early career consultation groups." These groups would travel to or virtually participate in some subset of decadal survey meetings. This would give early career astronomers a ``seat at the table" but would also mitigate the potentially deleterious effects of full survey participation on the careers of participants in these groups. 

Other ideas for early career involvement ranged from early career town halls, workshops, and conferences hosted by the American Astronomical Society and National Academies to inviting early career astronomers to make presentations to the decadal panels. While the assembled astronomers were not uniformly in support of a single recommendation, increased participation in some format (and perhaps multiple formats) was a universal theme in the discussions. 
\paragraph{Recommendations}
In order to increase the early career representation in the decadal survey process, we recommend that the decadal survey committee:
\begin{itemize}
    \item \textbf{2.1.1} Implement incentives for early career astronomers to participate as full panel members (see above).
    \item \textbf{2.1.2} Invite early career astronomers to serve as consultants to the decadal panels.
    \item \textbf{2.1.3} Invite early career astronomers to deliver presentations to panels.
\end{itemize}

\subsection{Soliciting inclusive input to the 2020 decadal and beyond} Another discussion theme revolved around how to solicit broad input from the astronomical community, ensure that diverse voices are heard, and successfully disseminate the conclusions of the decadal survey to all of its ``stakeholders." First, town halls could be standardized (so that more places could easily arrange them) and conducted both digitally and in a wide range of geographic locations. Focused virtual town halls could explicitly target the challenges facing certain groups in astronomy (e.g. postdocs, observatory staff). 

Second, we suggest that panel members are either present at the town halls, or that minutes from these events are compiled and circulated so that panel members are aware of what was discussed. Lastly, the survey committee should also consider inviting additional stakeholders not traditionally involved in the decadal survey process (e.g. native Hawaiians, amateur astronomers, K12 astronomy educators, data science employers) and certain additional subsets of the community (e.g. disabled astronomers, LGBTQ+ astronomers, etc.) to participate in Astro2020, either by making invited presentations to the committee or through focused town halls that are broadly advertised.   

\paragraph{Recommendations} 
In order to ensure equitable participation and dissemination to the broadest possible audience of stakeholders (within and beyond the astronomy community), we recommend that the decadal survey committee:
\begin{itemize}
    \item \textbf{2.2.1} Standardize town halls.
    \item \textbf{2.2.2} Organize official gatherings (town hall, focus session, etc.) that target specific groups of people whose perspectives have been historically missing in the decadal survey process and/or are important for the current decadal at hand (i.e. underrepresented people groups, geographically underrepresented groups, graduate students, postdocs, etc.).
    \item \textbf{2.2.3} Encourage participation by all stakeholders in the decadal survey process through invited presentations to the committee and targeted virtual town halls.
    \item \textbf{2.2.4} Make survey highlights available in multiple easily-digestible formats that adhere to the principles of universal design. For example: an ``Executive Summary" (1-page) document, publicly available slide decks highlighting priorities, digital representations that are captioned and easy to share on social media platforms (e.g. 5-10 minute videos). 
\end{itemize}

\section{Conclusion}
The Early Career Astronomer and Astrophysicist Focus Session was a unique opportunity for a large group of early career professionals to gather and discuss the state of the astronomy profession. Lively and at times contentious discussions about the state of the profession centered around the five main themes identified in this document, namely: the role of big data in the future of astronomy, the evolving challenges of large missions and collaborations, graduate training and career preparation, the postdoctoral years, and creating a diverse workforce and equitable community. Although ECFS participants had many concerns about and ideas for improvements in these areas, there was also an overwhelming energy and enthusiasm about the future of our field and an optimism about the potential for meaningful change. A central goal of the event was to collect ideas about changes that early career astronomers would like to see made to the decadal survey process, and ideas that were generated in this regard are outlined in section 2. We, the participants, felt strongly that early career voices should be heard at every stage of the decadal process, and not just at the focus session. We look forward to participating in more of the Astro2020 decadal survey. 






\end{document}